\documentclass[twocolumn,twoside,slac_two]{revtex4}
\usepackage{graphicx}
\usepackage{fancyhdr}
\def\be{\begin{equation}}
\def\ee{\end{equation}}
\def\ba{\begin{eqnarray}}
\def\ea{\end{eqnarray}}

\def\mpl{m_{\rm p}}
  \def\be{\begin{equation}}
\def\ee{\end{equation}}
 \def\bi{\begin{itemize}}
 \def\ei{\end{itemize}}
  \def\ben{\begin{enumerate}}
\def\een{\end{enumerate}}
  \def\bt{\begin{tabular}}
\def\et{\end{tabular}}
\def\bc{\begin{center}}
\def\ec{\end{center}}

\def\bea{\begin{eqnarray}}
\def\eea{\end{eqnarray}}

\begin{document}

\title{Interactions and Instabilities in Cosmology's Dark Sector}

%

\author{Mark Trodden}
\affiliation{Center for Particle Cosmology, Department of Physics and Astronomy, University of Pennsylvania,
Philadelphia PA 19104, USA.}

\begin{abstract}
I consider couplings between the dark energy and dark matter sectors. I describe how the existence of an adiabatic regime, in which the dark energy field instantaneously tracks the minimum of its effective potential, opens the door for a catastrophic instability. This {\it adiabatic instability} tightly constrains a wide class of interacting dark sector models. This talk was presented at, and will appear in the proceedings of the DPF-2009 conference.
\end{abstract}

\maketitle

\thispagestyle{fancy}


\section{Introduction}
Modern cosmology demands the existence of two new components to the energy budget of the universe. Dark matter is necessary for structure formation to occur properly, and to account for numerous observations, such as gravitational lensing, the comic microwave background (CMB) and galaxy rotation curves. The discovery of cosmic acceleration requires a second source of new physics, which may come in the form of a modification to general relativity, but which is perfectly consistent with a cosmological constant or a dynamical dark energy component~(for reviews see~\cite{Copeland:2006wr,Linder:2008pp,Frieman:2008sn,Silvestri:2009hh,Caldwell:2009ix}).

At a phenomenological level, this description is a remarkable fit to all current observations. However, at the level of fundamental physics the existence of dark matter and dark energy, comprising the vast majority of the contents of the universe, poses a critical challenge: how do these components fit into our microphysical theories of matter and energy? One way to search for such relationships is to explore observable consequences of interactions between dark components and baryonic matter. Examples of this are the search for the annihilation products of dark matter and collider searches for dark matter.

Alternatively, if dark matter and dark energy are to fit into a unified description, we might expect~\cite{combined} that there would also be interactions between them. In this talk I briefly explored some consequences of such interactions, focusing on a possible catastrophic instability - the {\it adiabatic instability} - and a number of constraints on stable models. As with all proceedings with page limits, thorough referencing is impossible, and I have therefore referenced mostly review articles, and the papers I actually referred to in the talk.

\section{Modeling Dark Couplings}
I'll focus on the following action, encapsulating many models studied in the
literature~\cite{combined1},
\bea
S[g_{ab},\phi,\Psi_{\rm j}]
&=& \int d^4x\sqrt{-g}
\left[ \frac{1}{2} \mpl^2 R
-\frac{1}{2} (\nabla \phi)^2
 - V(\phi)
\right] \nonumber \\
&&+ \Sigma_{\rm j} S_{\rm j}[e^{2 \alpha_{\rm j}(\phi)} g_{\mu\nu}, \Psi_{\rm j}] \ ,
\label{action0}
\eea
where $g_{\mu\nu}$ is the Einstein frame metric, $\phi$ is a scalar field
which acts as dark energy, and $\Psi_{\rm j}$ are the matter fields. The signature is $(-,+,+,+)$ and I 
define the reduced Planck mass by $\mpl^2 \equiv (8\pi G)^{-1}$.
The functions $\alpha_{\rm j}(\phi)$ are couplings to the j${}^{th}$ matter sector.

The field equations following from this are
\bea
\mpl^2 G_{\mu\nu} &=& \nabla_\mu \phi \nabla_\nu \phi - \frac{1}{2} g_{\mu\nu}
(\nabla \phi)^2 - V(\phi) g_{\mu\nu} \nonumber \\
&& + \sum_{\rm j} e^{4 \alpha_{\rm j}(\phi)}
\left[ ({\bar \rho}_{\rm j} + {\bar p}_{\rm j}) u_{{\rm j}\,\mu} u_{{\rm j}\,\nu} + {\bar p}_{\rm j} g_{\mu\nu} \right] \ ,
\label{ee0d}
\eea
\be
\nabla_\mu \nabla^\mu \phi - V'(\phi) = \sum_{\rm j} \alpha_{\rm j}'(\phi) e^{4 \alpha_{\rm j}(\phi)}
({\bar \rho}_{\rm j} - 3 {\bar p}_{\rm j} ) \ ,
\label{eq:scalar10a}
\ee
where I have treated the matter field(s) in the j${}^{th}$ sector as a fluid with density
${\bar \rho}_{\rm j}$ and pressure ${\bar p}_{\rm j}$ as measured in
the frame $e^{2 \alpha_{\rm j}} g_{\mu\nu}$, and with
4-velocity $u_{{\rm j}\,\mu}$ normalized according to $g^{\mu\nu} u_{{\rm j}\,\mu} u_{{\rm j}\,\nu}=-1$.

For simplicity, let us neglect baryons, and consider a
composite dark matter sector, with one coupled species with density
$\rho_c$ and coupling $\alpha_c(\phi) = \alpha(\phi)$, and another
uncoupled species with density $\rho_{co}$ and coupling $\alpha_{co}=0$.
Since these both represent types of dark matter ${\bar p}_c =
{\bar p}_{co}=0$, and it is convenient to set $\rho_{\rm j} =
e^{3 \alpha_{\rm j}} {\bar \rho}_{\rm j}$ to yield
\bea
\mpl^2 G_{\mu\nu} &=& \nabla_\mu \phi \nabla_\nu \phi - \frac{1}{2} g_{\mu\nu}
(\nabla \phi)^2 - V(\phi) g_{\mu\nu}\nonumber
\\ && + e^{\alpha(\phi)}\rho_c u_{c\mu} u_{c\nu}+ \rho_{co} u_{co\mu} u_{co\nu} \ ,
\label{ee}
\eea
and $
\nabla_\mu \nabla^\mu \phi - V_{\rm eff}'(\phi) = 0.
$
Here the effective potential is given by
$
V_{\rm eff}(\phi) = V(\phi) + e^{\alpha(\phi)}\rho_c \ ,
$
and the fluid obeys $\nabla_\mu ( \rho_c u_c^\mu) =0$,
and $u_c^\nu \nabla_\nu u_c^\mu  = - (g^{\mu\nu} + u_c^\mu u_c^\nu) \nabla_\nu \alpha$.

\section{The Adiabatic Regime}
The effective potential $V_{\rm eff}(\phi)$ may have a minimum resulting from the
competition between the two distinct terms. If the timescale or lengthscale for $\phi$ to adjust
to the changing position of the minimum of $V_{\rm eff}$ is shorter
than that over which the background
density changes, the field $\phi$ will adiabatically track
this minimum~\cite{Das:2005yj}.

\begin{figure}[t]
\begin{center}
\includegraphics[width=3in]{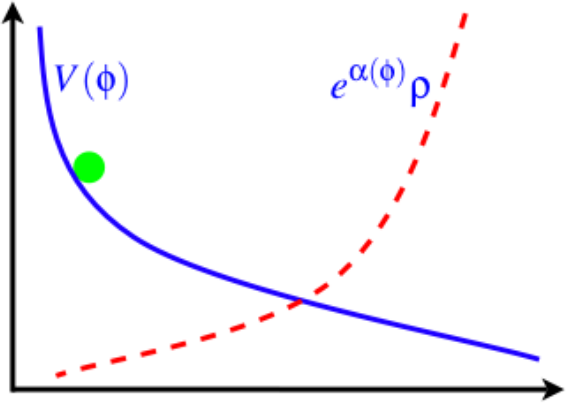}
\caption{The effective potential felt by the dark energy field has contributions from the raw potential $V(\phi)$ and from the nontrivial coupling $\alpha(\phi)$ of $\phi$ to the background dark matter density $\rho$. The resulting minimum allows for the possibility of an adiabatic regime.}
\end{center}
\end{figure}
In this case the coupled CDM component with $\phi$ together
acts as a single fluid
with an effective energy density
\be
\rho_{\rm eff}(\rho_c) = e^{\alpha[\phi_{\rm m}(\rho_c)]} \rho
+ V[\phi_{\rm m}(\rho_c)] \ ,
\label{rhoeff}
\ee
and effective pressure
\be
p_{\rm eff}(\rho_c) = - V[\phi_{\rm m}(\rho_c)] \ .
\label{peff}
\ee
Here $\phi_{\rm m}(\rho_c)$ is the solution of the algebraic
equation
\be
V_{\rm eff}'(\phi)= V'(\phi) + \alpha'(\phi) e^{\alpha(\phi)} \rho_c =0
\label{eq:alg}
\ee
for $\phi$.
Eliminating $\rho_c$ between Eqs.\ (\ref{rhoeff}) and (\ref{peff}) gives
the equation of state $p_{\rm eff} = p_{\rm eff}(\rho_{\rm eff})$.

The coupled fluid acts as the
source of cosmic acceleration, and in the adiabatic approximation the effective fluid description is valid
for the background cosmology and for linear and nonlinear perturbations. Therefore,
the equation of state of perturbations is the same as that of the background cosmology, and
the matter and scalar field evolve as one effective fluid,
obeying the usual fluid equations of motion with the given effective
equation of state.

A necessary condition for the validity of the adiabatic approximation is that the
lengthscales or timescales ${\cal L}$ over which the density $\rho_c$ varies
are large compared to inverse of the effective mass of the scalar field.  A more precise condition can be shown to be~\cite{Bean:2007ny}
\be
\frac{ d \ln V[\phi_{\rm m}(\rho_c)]}{d \ln \rho_c} \left(
  \frac{1}{m_{\rm eff}^2 {\cal L}^2} \right) \ll 1\ .
\label{condt1}
\ee
This condition justifies dropping terms
involving the gradient of $\phi$ from the fluid and Einstein
equations. Many dark energy models admit regimes in which this condition is satisfied for the background and for
linearized perturbations over a range of scales.

In the adiabatic regime, the
inferred dark energy equation of state parameter (neglecting baryons) is
\be
w = \frac{-1}{1 - (1-e^{(\alpha_0 -\alpha)} ) \frac{d \ln V }{d \alpha} } \ ,
\ee
with $\alpha_0\equiv\alpha(\phi_0)$ the value today. Thus, $w$ is precisely $-1$ today,
and generically satisfies $w < -1$ in the past~\cite{Das:2005yj,Bean:2007ny}


\section{The Adiabatic Instability}
The
potential $V(\phi)$ may be written as a function $V(\alpha)$ of the
coupling function $\alpha(\phi)$ by eliminating $\phi$.
This gives, from Eqs.\ (\ref{rhoeff}) and (\ref{eq:alg}),
\be
\rho_{\rm eff} = V + e^{ \alpha} \rho_c = V - \frac{
  dV/d\phi}{d\alpha/d\phi} = V - \frac{dV}{d\alpha} \ .
\ee
The
square of the adiabatic sound speed, $c_{a}^{2}=\dot{P}/\dot\rho$ is then given by
\be
\frac{1}{c_a^2} = \frac{d \rho_{\rm eff}}{dp_{\rm eff}}
= \frac{ d\rho_{\rm eff} / d\alpha}{d p_{\rm eff} / d\alpha}
= -1 + \frac{ \frac{d^2 V}{d\alpha^2} }{
  \frac{dV}{d\alpha}} \ .
\label{soundspeed}
\ee
In the adiabatic regime the effective sound speed relevant for local perturbations in pressure and density, $c_s^2(k,a)\equiv \delta P(k,a)/\delta\rho(k,a)$, tends towards the adiabatic sound speed and is {\it always negative}, since
$dV/d\alpha$ must be negative so that Eq.\ (\ref{eq:alg}) admits a solution,
and $d^2V/d\alpha^2$ must be positive to ensure a
positive $m_{\rm eff}^2$. 

Consider  the regime in which this adiabatic limit has been reached ($c_s^2=c_a^2$) and focus on
a perturbation with lengthscale ${\cal L}$.
In order to be in the adiabatic regime we require ${\cal L} \gg m_{\rm
 eff}^{-1}$.  The negative sound speed squared will cause an
exponential growth of the mode, as long as the
growth timescale $\sim {\cal L} / \sqrt{|c_s^2|}$ is short compared to
the local gravitational timescale $\mpl /\sqrt{\rho_{\rm eff}(\rho_c)}$.
SInce $c_s^2 m_{\rm eff}^2 = (\alpha')^2 V_{,\alpha} = (\alpha')^2
\rho_{\rm eff} / (V/V_{,\alpha}-1)$,
the instability will operate in the range of lengthscales given by
\be
\frac{1}{m_{\rm eff}(\rho_c)} \ll  {\cal L} \ll \frac{\mpl
|\alpha^\prime[\phi_{\rm m}(\rho_c)]|}{m_{\rm eff}(\rho_c)}
\sqrt{\frac{1}{1 - \frac{1}{\frac{d\ln V}{d \alpha}}}}.
\label{range}
\ee
Here the quantity $d \ln V / d\alpha(\alpha)$
on the right hand side is
expressed as a function of $\phi$ using $\alpha = \alpha(\phi)$, and
then as a function of the density using $\phi = \phi_{\rm m}(\rho_c)$.
In order for this range of scales to  be non empty, the dimensionless
coupling $\mpl |\alpha'|$ must be large compared to unity, i.e., the
scalar mediated interaction between the dark matter particles must be
strong compared to gravity (see also~\cite{Mota:2006ed,Mota:2006fz}.

\subsection{Understanding the Instability}
There are two different ways of describing and understanding the instability, depending
on whether one thinks of the
scalar-field mediated forces as ``gravitational'' or ``pressure'' forces.

In the Einstein frame, the
instability is independent of gravity, since it is present even when
the metric perturbation due to the fluid can be
neglected.  In the adiabatic regime the acceleration due the scalar
field is a gradient of a local function of the density, which can be thought of as a pressure.  The net
effect of the scalar interaction is to give a contribution to the
specific enthalpy $h(\rho_c) = \int dp/\rho_c$ of any fluid which is independent
of the composition of the fluid.
If the net sound speed squared of the
fluid is negative, then there exists an instability in accord with our
usual hydrodynamic intuition.

On the other hand, in the Jordan
frame description, the instability most certainly involves gravity.
The effective Newton's constant describing the interaction of dark matter with itself is
\be
G_{cc} = G \left[ 1
  + \frac{2 \mpl^2 \alpha^\prime(\phi)^2 }{1 +
\frac{m_{\rm eff}^2}{ {\bf k}^2 }
} \right],
\label{Gformula0}
\ee
where ${\bf k}$ is a spatial wavevector \cite{Bean:2007ny}.
At long lengthscales the scalar interaction is
suppressed and $G_{cc} \approx G$.  At short lengthscales,
the scalar field is effectively massless and
$G_{cc}$ asymptotes to a constant.
However, when $\mpl |\alpha^\prime| \gg 1$ there is an intermediate range
\be
m_{\rm eff}(\mpl |\alpha^\prime|)^{-1} \ll k \ll m_{\rm eff}
\label{range3}
\ee
over which the effective Newton's constant increases like $G_{cc} \propto {\bf k}^2$.
This interaction behaves just
like a (negative) pressure in the hydrodynamic equations.
This
explains why the the effect of the scalar interaction can be thought
of as either pressure or gravity in the range of scales (\ref{range3}).
Note that the range of scales (\ref{range3}) coincides with with the
range (\ref{range}) derived above, up to a logarithmic correction factor.

From this second, Jordan-frame point of view, the instability is
simply a Jeans instability.  In a cosmological background
the CDM fractional density perturbation traditionally exhibits power-law growth on subhorizon scales because Hubble damping competes with the exponential (Jeans) instability one might expect on a timescale of $1/ \sqrt{G \rho}$.
In our case, however, the gravitational self-interaction of the mode is governed by $G_{cc}(k)$ instead of $G$, and consequently in the range (\ref{range3}) where $G_{cc} \gg G$ the timescale for the Jeans instability is much shorter than the Hubble damping time.
Therefore the Hubble damping is ineffective and the Jeans instability
causes approximately exponential growth.



\section{Examples of Theories with an Adiabatic Instability}
\label{examples}
\noindent
{\it 1. Exponential Potential and Constant Coupling}
\\

The canonical example is a theory with an exponential potential of the form
\be
V = V_0 e^{-\lambda \phi/\mpl} \ ,
\label{expV}
\ee
with $\lambda>0$ and with linear coupling functions
\be
\alpha(\phi) = - \beta C \frac{\phi}{\mpl} \ ,
\label{constantcoupling}
\ee
where $\beta = \sqrt{2/3}$ and $C$ is a constant.
The effective potential is
\be
V_{\rm eff}(\phi,\rho) = V_0 e^{-\lambda \phi/\mpl} + e^{-\beta C \phi/\mpl} \rho \ ,
\ee
and solving for the local minimum of this potential yields
the relation between $\phi$ and $\rho$ in the adiabatic regime:
\be
e^{(\lambda -\beta C) \phi_{\rm m}(\rho)/\mpl} = \frac{ \lambda V_0 }{-\beta C
  \rho} \ .
\label{phimans}
\ee
Note that $C$ must be negative in order for
the effective potential to have a local minimum and for an adiabatic
regime to exist.  Restricting attention to this case, and
defining the dimensionless
positive parameter $\gamma = - \lambda / \beta C$,
the corresponding effective mass parameter is
\be
m_{\rm eff}^2 = \lambda^2 \mpl^{-2} V_0 \frac{1+\gamma}{\gamma} \left( \frac{
    \rho}{\gamma V_0} \right)^{\frac{\gamma}{\gamma+1}} \ .
\label{mass1}
\ee

Using~(\ref{soundspeed}) we obtain the sound speed squared as
\be
c_s^2  =  - \frac{1}{1+\gamma} \ ,
\label{cs21}
\ee
so this
model is always unstable in the adiabatic regime.
Eqs.\ (\ref{expV}) and (\ref{phimans}) also yield
\be
\frac{\partial \ln V}{\partial \ln \rho} = \frac{\gamma}{1 + \gamma} \ ,
\label{logd}
\ee
which allows one to calculate the range of spatial scales 
${\cal L}_{\rm min}(\rho) \ll {\cal L} \ll {\cal L}_{\rm max}(\rho)$ over which
the instability operates for a given density $\rho$, where
\be
{\cal L}_{\rm min}(\rho)^2 = \frac{\gamma^2 }{\lambda^2 (1 + \gamma)^2 }
\frac{\mpl^2}{V_0}
\left( \frac{ \gamma
    V_0}{ \rho} \right)^{\frac{\gamma}{\gamma+1}}
\label{Lmin2}
\ee
and
\be
{\cal L}_{\rm max}(\rho)^2 = \beta^2 C^2 {\cal L}_{\rm min}(\rho)^2 \ .
\label{Lmax2}
\ee
Thus, there is a nonempty unstable regime only when $\beta |C| \gg 1$,
ie with the scalar coupling is strong compared to the gravitational
coupling, as we saw earlier.

To see the effect of the instability more explicitly, consider cosmological perturbations.
The Einstein-frame FRW equation in the adiabatic limit is
\be
3 \mpl^2 H^2 = V + e^\alpha \rho \ ,
\label{cosmo1}
\ee
where $\rho \propto 1/a^3$.  This yields $a(t) \propto t^{2/(3 + 3 w_{\rm eff})}$, where
the effective equation of state parameter is
\be
w_{\rm eff} = - \frac{1}{1 + \gamma} \ .
\label{weff1}
\ee
In the strong coupling limit $|C| \to \infty$, $w_{\rm eff} \to -1$.
Thus the adiabatic regime of this model with large $|C|$ is incompatible with
observations in the matter dominated era, where $w_{\rm eff} \approx
0$ except for at small redshifts.
Nevertheless, the model is still
useful as an illustration of the instability.

From~(\ref{Lmin2}), (\ref{Lmax2}) and (\ref{cosmo1})
the range of unstable scales is given by
\be
\frac{1}{\beta^2 C^2} \ll \frac{H^2 a^2}{k^2}  \ll \frac{1}{3(1+\gamma)} \ ,
\ee
where $k$ is comoving wavenumber.  This range of scales always lies
just inside the horizon.  A given mode $k$ will evolve through this
unstable region before it exits the horizon.

One can then show that the
perturbation evolution equation, specialized to the exponential model, and in the strong coupling limit
$|C| \to \infty$ is
\be
\frac{d^2 \delta}{d a^2} + \frac{3}{a} \frac{d \delta}{da} -
\frac{k^2}{ H^2 a^4} \delta =0 \ .
\ee
In the strong coupling limit $H$ is approximately a constant, $H
\approx H_0$, and the growing mode solution is
\be
\delta(a) \propto \frac{1}{a} K_1\left(\frac{k}{H_0 a}\right) \approx
\sqrt{ \frac{\pi H_0}{2 k a}} \exp \left(-\frac{k}{H_0 a} \right) \ ,
\ee
where $K_1$ is the modified Bessel function.  The mode grows by a
factor $\sim e$ when the scale factor changes from $a$ to $a + \Delta
a$, where $\Delta a/a \sim a H_0 / k \ll 1$ for subhorizon
modes.

A more detailed analysis of the cosmology of this model is given in~\cite{Bean:2008ac}, but in the non-adiabatic regime $|C| \sim 1$ rather than the
strong coupling regime $|C| \gg 1$ considered here.
\vspace{5mm}

\noindent
{\it 2.Two Component Dark Matter Models}
\\
As a more realistic example, consider models in which there are two dark matter sectors, a
density $\rho_c$ which is not coupled to the scalar field, and a
density $\rho_{co}$ which is coupled with coupling function (\ref{constantcoupling})
and exponential potential (\ref{expV}).
Both of these components are treated as pressureless fluids.
The FRW equation for this model in the adiabatic limit is
\be
3 \mpl^2 H^2 = V + e^\alpha \rho_{co} + \rho_c \ .
\label{cosmo2}
\ee
The first two terms on the right hand
side of Eq.\ (\ref{cosmo2}) act like a fluid with equation of state
parameter given by (\ref{weff1}), and in the strong coupling limit
$|C| \gg 1$ this
fluid acts like a cosmological constant.  Thus, the background
cosmology can be made close to $\Lambda$CDM by taking $|C|$ to be
large.

The fraction of dark matter which is coupled must be small in the
limit of large coupling, $|C| \gg 1$.
Denoting $\Omega_V = V / (3 \mpl^2 H^2)$,
$\Omega_{co} = e^\alpha \rho_{co} / (3 \mpl^2 H^2)$ and
$\Omega_c = \rho_c / (3 \mpl^2 H^2)$ gives $1 = \Omega_V +
\Omega_{co} + \Omega_c$.  Also from Eq.\ (\ref{phimans}) it follows
that, if the asymptotic adiabatic regime  has been reached, $\Omega_{co} = \gamma \Omega_V$, yielding
\be
\Omega_{co} = \frac{\gamma}{1 + \gamma} (1 - \Omega_c) \ .
\label{Omegaco}
\ee
Since $\Omega_c \sim 0.3$ today, and $\gamma \ll 1$ in the strong
coupling limit, we must have $\Omega_{co} \ll 1$
today.

The maximum and minimum lengthscales for the instability are still
given by Eqs.\ (\ref{Lmin2}) and (\ref{Lmax2}), but with $\rho$ replaced by
$\rho_{co}$.  Since $\rho_{\rm co}$ is approximately a constant in the
strong coupling limit, these lengthscales are also constants.
If the parameters of the model are chosen so that $\Omega_c \sim 1$
today, then
\be
{\cal L}_{\rm max} \sim H_0^{-1}, \ \ \ \ {\cal L}_{\rm min} \sim
\frac{H_0^{-1}}{\beta |C|} \ .
\label{scales0}
\ee
The evolution equations for the fractional density perturbations
$\delta_{\rm j} = \delta \rho_{\rm j} / \rho_{\rm j}$
in the adiabatic limit on subhorizon scales
are given by
\bea
{\ddot \delta}_c + 2 H {\dot \delta}_c &=& \frac{1}{2 \mpl^2} \rho_c \delta_c +
\frac{1}{2 \mpl^2} e^\alpha \rho_{co} \delta_{co}, \\
{\ddot \delta}_{co} + 2 H {\dot \delta}_{co} &=& \frac{1}{2 \mpl^2}
\rho_c \delta_c \nonumber \\
&&+
\frac{1}{2 \mpl^2} \left[ 1 + \frac{2 \beta^2 C^2}{1 + \frac{m_{\rm
        eff}^2 a^2}{k^2}} \right] e^\alpha \rho_{co}
\delta_{co} \ . \nonumber \\
\label{uns}
\eea
The condition for the instability to operate is that
the timescale associated with the second
term on the right hand side of Eq.\ (\ref{uns}) be short compared with
$H^{-1}$, or
\be
\frac{\beta^2 C^2 k^2}{m_{\rm eff}^2 a^2} \rho_{co} e^\alpha \gg H^2
\mpl^2 \ .
\label{condt4}
\ee
Now the effective mass for this model is given by
$m_{\rm eff}^2 = \beta^2 C^2 \mpl^{-2} (V + e^\alpha \rho_{co}) = 3
\beta^2 C^2 \mpl^{-2} H^2 (\Omega_V + \Omega_{co})$.  Substituting this
into Eq.\ (\ref{condt4}) and using Eq.\ (\ref{Omegaco}) gives the criterion
$k/(a H) \gg 1$.  Therefore
the instability should operate whenever modes are inside the horizon
and in the range of scales (\ref{scales0}).

\begin{figure}[t]
\begin{center}
\includegraphics[width=3.5in]{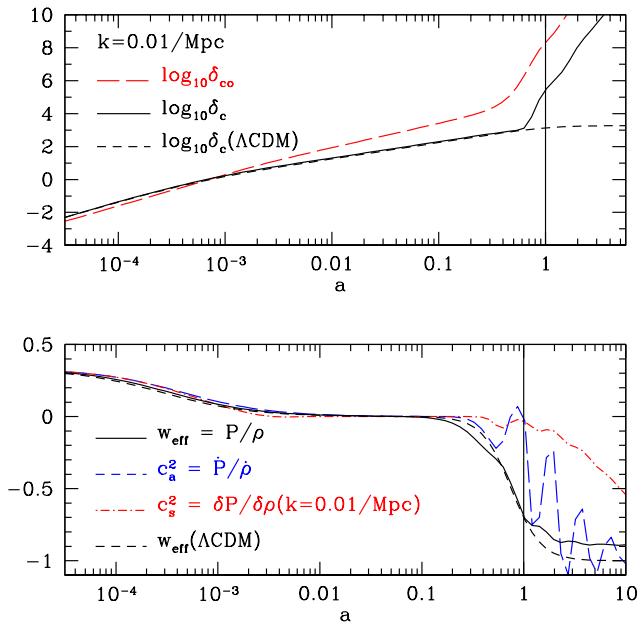}
\caption{[Bottom] The two component coupled dark energy (CDE)
  model. At late times the scalar field finds the adiabatic minimum with asymptotic equation of state, and sound speed $= -1/(1+\gamma) = -0.89$, able to reproduce a viable background evolution consistent   with supernovae, CMB angular diameter distance and BBN expansion
  history constraints. The figure shows the evolution of the effective equation of state, $w_{eff}$ (black full line), the adiabatic speed of sound, $c_{a}^{2}=\dot{P}/\dot{\rho}$ for all components (blue long dashed line) and for the coupled components only (green dot long dashed line), and effective speed of sound for $c_s^2=\delta P/\delta\rho$ at $k=0.01/Mpc$ for all components (red dot-dashed line)  and for the coupled components alone (magenta dotted line). The effective equation of state for a comparable $\Lambda$CDM model with $\Omega_{c}=0.25$, $\Omega_b=0.05$ and $\Omega_{\Lambda}=0.7$ is also shown (black dashed line). [Top] The growth of the fractional over-density $\delta=\delta\rho/\rho$ for $k=0.01/Mpc$ for the coupled CDM component, $\delta_{co}$, (red long dashed line) and uncoupled component, $\delta_{c}$, (black full line) in comparison to the growth for the $\Lambda$CDM model (black dashed line).
  \label{fig1}} 
\end{center}
\end{figure}

These expectations are confirmed by numerical integrations.
Figure \ref{fig1}  shows a numerical analysis
of such a two component model with an exponential potential with $\lambda =2$, strong
coupling ($C=-20$), and typical cosmological parameters,
$H_0=70 \, {\rm km} \, {\rm s}^{-1}\,{\rm Mpc}^{-1}$,
$\Omega_{b}=0.05$, $\Omega_{c}=0.2$, $\Omega_{co}=0.05$, and
$\Omega_{V}=0.7$. The existence of a dynamical attractor renders
the final evolution largely insensitive to the initial conditions for $\phi$, and the effect of the coupled CDM component peculiar velocity in the initial conditions can be neglected, since it is many orders of magnitude smaller than  the density perturbation. The
background evolution is entirely consistent with a $\Lambda$CDM like
scenario. The large
coupling drives the evolution to an adiabatic regime at late times,
with an adiabatic sound speed $c_{a}^{2}\rightarrow -1/(1+\gamma)$ as
in (\ref{cs21}). This drives a rapid growth in over-densities once in the adiabatic regime so that although consistent with structure observations at early times, they are inconsistent once the accelerative regime has begun.

In summary, these models provide a class of theories for which the
background cosmology is compatible with observations, but which are
ruled out by the adiabatic instability of the perturbations.

\section{Conclusions}
I have briefly discussed a broad class of models in which dark matter is coupled to a dark energy component, assumed to be responsible for the acceleration of the universe. Within these models, I have discussed a possible instability - the adiabatic instability - that arises in a range of cosmological and astrophysical settings, and which rules out a set of parameter values. I have discussed specific examples of models in which this instability is active, and it is worth noting that one may carry out similar analyses to constrain subclasses of chameleon and MaVan models.


\begin{acknowledgments}
I would like to thank Rachel Bean and Eanna Flanagan for enjoyable collaborations and permission to use figures from our joint work here. I would also like to thank the organizers of DPF 2009 for all their hard work in Detroit. This work was supported in part by the National Science Foundation under grant PHY-0930521, by Department of Energy grant DE-FG05-95ER40893-A020 and by NASA ATP grant NNX08AH27G.
\end{acknowledgments}

\bigskip 

\end{document}